\newcommand{\beq}{\begin{equation}}
\newcommand{\eeq}{\end{equation}}
\newcommand{\bea}{\begin{eqnarray}}
\newcommand{\eea}{\end{eqnarray}}
\newcommand{\bear}{\begin{eqnarray*}}
\newcommand{\eear}{\end{eqnarray*}}
\newcommand{\scp}{\scriptsize }

\newcommand{\rf}[1]{(\ref{#1})}

\documentclass[pra,preprint]{revtex4}

\usepackage{psfrag}
\usepackage{graphicx}

\begin{document}
\title
{Phase diagram and spectral properties of a new exactly integrable spin one quantum chain}
\author{
 Francisco C. Alcaraz and Gilberto M. Nakamura}
\affiliation{Instituto de F\'{\i}sica de S\~ao Carlos,\\
Universidade de S\~ao Paulo, 
CP 369, 13560-970, S\~ao Carlos, SP, Brazil}, 
\date{\today}

\begin{abstract}

The spectral  properties and phase diagram of the 
exact integrable spin one quantum chain introduced by Alcaraz 
and Bariev are  presented. The model has a $U(1)$ symmetry and  its 
integrability is associated to an unknown R-matrix whose dependence 
on the spectral 
 parameters is not of difference form. The associated Bethe ansatz equations, that fix the eigenspectra, are distinct from  those associated to other known integrable spin models. The model has a free parameter $t_p$. We show that at the special 
point $t_p=1$ the model gets an extra $U(1)$ symmetry and 
 reduces to the deformed $SU(3)$ Perk-Schultz 
model at a special value of its anisotropy $q=\exp(i2\pi/3)$ and 
in the presence of an external magnetic field. 
 Our analysis is done either by solving the associated 
Bethe-ansatz equations or by direct diagonalization of the 
quantum Hamiltonian for small lattice sizes. The phase diagram is calculated by exploring the consequences of conformal invariance on the finite-size corrections of the Hamiltonian eigenspectrum. The model exhibits a critical phase ruled by $c=1$ conformal field theory separated from a massive phase by   first-order  phase transitions.

\end{abstract}

\maketitle

\section{Introduction}

The anisotropic spin $1/2$ Heisenberg model, or XXZ quantum 
chain, and the 6-vertex model are considered as paradigm of exact integrability in statistical mechanics \cite{yang}. In 
the XXZ quantum chain the $z$-component of the total 
magnetization is a good quantum number ($U(1)$ symmetry). 
Its simplest integrable generalizations that keeps the 
$U(1)$ symmetry are spin-1 quantum chains. Models on this 
class are the Fateev-Zamolodchikov model \cite{fateev}, the 
Izergin-Korepin model \cite{izergin}, the supersymmetric 
$OSP(1/2)$ model \cite{virchirko} and the biquadratic model 
\cite{biquadratic}. The integrability of these models is a 
consequence of the existence of a known associated R-matrix 
satisfying the Yang-Baxter equation. The associated R-matrix 
for these models  
are regular, i.e., depends only on the difference of the 
spectral parameters. 

A new exact integrable spin-1 quantum chain was derived by 
using  the coordinate 
Bethe ansatz \cite{AB1}, or a matrix product ansatz 
\cite{MPA}.
 The derivation  of the integrable model through these last 
approaches does not depend on the knowledge of the associated $R$ matrix. 
Distinct from the other  integrable \cite{fateev}-\cite{biquadratic} 
and non integrable \cite{blume} spin-1 models, whose physical 
properties are well studied, almost no physical information 
is known for this new quantum chain, besides its exact 
integrability. The unknown associated R-matrix  is not 
regular \cite{AB1} since it does not satisfy the Reshetikihin 
criterion \cite{kulish}. The Bethe ansatz equations (BAE) that fix the eigenenergies are also quite distinct from the 
corresponding equations of other spin-1  integrable quantum 
chains. 

In this paper we are going to present an extensive analytical 
and numerical analysis of the eigenspectra properties  of 
this new spin-1 quantum chain. Based on solutions of the 
associated BAE,
whenever it is possible, and diagonalizations of the quantum 
Hamiltonian  
 on small lattices ($L=2-24$) we are able to predict some of  its 
critical properties. 

The paper is organized as follows. In section 2 we present 
the model and the BAE that fix the eigenspectra.  
The model has a $U(1)$ symmetry and is exact integrable for 
any value of a free parameter $t_p$. We show, that
 for the special value $t_p =1$, 
the model is 
related to the deformed $SU(3)$ Perk-Schultz model with 
deformation parameter
$q=\exp(i2\pi/3)$ \cite{perk} in the presence of an external 
magnetic field.  In section 3 we 
analyze  the eigenspectra  of the quantum chain in 
several regions with distinct values of  the free 
parameter $t_p$. 
Based on 
conformal invariance predictions the critical properties of 
the model are  calculated.  Finally  in section 4 we summarize our results and present our conclusions.

\section{ The model} \label{sect2}

Instead of presenting the quantum Hamiltonian in terms of spin-1 $SU(2)$ 
matrices ($S^x,S^y,S^z$)  it is more convenient to present it in terms of the $3\times 3$ Weyl 
matrices $E^{l,m}$ ($l,m=0,1,2$), with elements $E^{l,m})_{i,j} = 
\delta_{l,i}\delta_{m,j}$. At each lattice site $i$ we may have zero particle 
($n_i=0$), one particle ($n_i=1$) or two particles ($n_i=2$) 
or equivalently $S_i^z =-1$, $S_i^z=0$ and $S_i^z=1$, respectively. The 
dynamics of these particles, in a periodic chain with $L$ sites, is 
described by the Hamiltonian
\beq \label{H1}
H(t_p,h) = - \sum_{j=1}^L \sum_{\alpha,\beta,\gamma,\mu =0}^2 
\Gamma_{\gamma,\mu}^{\alpha,\beta} 
E_j^{\gamma,\alpha}E_{j+1}^{\mu,\beta} 
-h\sum_{j=1}^L\sum_{\alpha=1}^2 \alpha E_j^{\alpha,\alpha}
\eeq
where
\bea \label{H2}
&&\Gamma_{0,1}^{1,0} = \Gamma_{1,0}^{0,1} = \Gamma_{1,2}^{2,1}=
\Gamma_{2,1}^{1,2} = -1, \quad 
\Gamma_{2,0}^{0,2}=\Gamma_{0,2}^{2,0} = -t_p, \nonumber \\
&& \Gamma_{0,2}^{1,1} =\Gamma_{1,1}^{0,2}= e^{-i\pi/3}\sqrt{t_p^2-1}, 
\quad \Gamma_{2,0}^{1,1}=\Gamma_{1,1}^{2,0} = -e^{i\pi/3}
\sqrt{t_p^2 -1},
\eea
are the hopping parameters (off diagonal) and the static terms (diagonal)
  are 
given by
\bea \label{H3}
&&\Gamma_{0,0}^{0,0} =
\Gamma_{2,2}^{2,2} = 0, \quad \Gamma_{1,0}^{1,0}=\Gamma_{0,1}^{0,1}=\frac{1}{4t_p}, \nonumber \\
&& \Gamma_{1,1}^{1,1} = t_p +\frac{1}{2t_p}, \quad \Gamma_{0,2}^{0,2} = 
\Gamma_{2,0}^{2,0} = -\frac{t_p}{2}, \nonumber \\
&& \Gamma_{1,2}^{1,2} = \frac{1}{4t_p} +i\frac{\sqrt{3}}{2}t_p, \quad 
\Gamma_{2,1}^{2,1} = \frac{1}{4t_p}-i\frac{\sqrt{3}}{2}t_p. 
\eea
The parameter $t_p$ is  free  and $h$ plays the role of a magnetic field in the $z$ direction ($z$-magnetic field)   or a chemical 
potential controlling the magnetization  or the number of particles in the ground state, respectively. 

The Hamiltonian \rf{H1} is non hermitian. 
It is interesting to observe that its non hermiticity  is
not only due to the presence of complex matrix elements in the diagonal 
 (see \rf{H3}) (as happens in the quantum deformed $SU_q(N)$ models) 
but also due to the presence of complex non diagonal elements (see \rf{H2}).
The Hamiltonian \rf{H1} has a $U(1)$ symmetry due to its commutation 
with the total density $\rho$ of particles:
\beq \label{H4}
\rho = \frac{n}{L},\quad n = \sum_{j=1}^L n_j, \quad n_j =  S_j^z +1
= \sum_{\alpha=1}^2 \alpha E_j^{\alpha,\alpha}.
\eeq
As a consequence its associated eigenvector space can be separated into 
disjoint sectors labeled by the the total number of particles $n$ (or 
density $\rho$) or equivalently by its magnetization. 

The quantum chain \rf{H1} corresponds to one of the exactly integrable 
models introduced in \cite{AB1}. It is given by the choice 
$\epsilon=1$ in Eq. (15) of \cite{AB1}. As compared with the original 
presentation of the model,  we also added 
a harmless $z$-magnetic field so that the ground state of \rf{H1} 
at $h=0$ has, for any value of $t_p$  total density $\rho=1$ or 
equivalently zero magnetization.  The Hamiltonian \rf{H1} is exact integrable for 
arbitrary values of the  parameter  $t_p$  and magnetic field 
$h$.  

 At $t_p=1$ the non-diagonal couplings $\Gamma_{2,0}^{1,1} =\Gamma_{1,1}^{2,0} = \Gamma_{0,2}^{1,1}=\Gamma_{1,1}^{0,2} = 0$ and the model has an 
additional $U(1)$ symmetry.  The number of sites with single and double occupancy  
are now conserved separately. 
The parameter $t_p$ can be interpreted as  an anisotropy parameter, being 
$t_p=1$ the isotropic point.
At this isotropic point, 
  apart from a contribution $i\frac{\sqrt{3}}{2}
\sum_{i=1}^L (E_i^{0,0}-E_{i+1}^{0,0})$, that vanishes in the 
periodic chain, 
 the Hamiltonian, with $h=0$, is given by 

\beq \label{H5}
H(t_p=1,h=0) = H_{PS}(i\frac{2\pi}{3}) -\frac{3}{2} 
\sum_{j=1}^{L}E_j^{1,1} + \frac{L}{2},
\eeq
where 
\bea \label{H6}
&& H_{PS}(\gamma) = \sum_{j=1}^L\sum_{\alpha=0}^2 \left\{\cosh \gamma 
E_i^{\alpha,\alpha}E_{i+1}^{\alpha,\alpha} \right.\nonumber \\
&&\left. + \sum_{\beta= \alpha+1}^2\left[\sinh \gamma 
(E_i^{\beta,\beta}E_{i+1}^{\alpha,\alpha} - 
E_i^{\alpha,\alpha}E_{i+1}^{\beta,\beta}) + 
E_i^{\alpha,\beta}E_{i+1}^{\beta,\alpha} + 
E_i^{\beta,\alpha}E_{i+1}^{\alpha,\beta}\right] \right\}
\eea
is the deformed spin-1 $SU(3)$ Perk-Schultz model \cite{perk} at the 
special value of the deformation parameter $q=e^{\gamma}$, $\gamma = 
i2\pi/3$. This model is also known as the anisotropic $SU(3)$ 
Sutherland model \cite{sut}. It is important to stress that the related  
Perk-Schultz Hamiltonian is the ferromagnetic one (signal $+$) in the 
presence of a special magnetic field (value $h^1=3/2$, $h^2 =0$) 
favoring single 
occupied sites. A simple calculation show us that the ground state 
for the related Perk-Schultz model happens in the sector with total 
density $\rho=1$. It corresponds to the trivial state $|11 \cdots 1>$  
where all the sites are single occupied. 
The model \rf{H1}, for $t_p \neq 1$,
 can then be 
considered as the spin-1 anisotropic Perk-Schultz model  at  $q=e^{i2\pi/3}$ 
with an additional parameter that breaks partially its symmetry. For 
arbitrary values of $t_p$ the eigenenergies and momentum are given by 
\cite{AB1}
\beq \label{H7}
E = -n(\frac{1}{2t_p} +h) + 2\sum_{j=1}^n \cos k_j, 
\quad P = \sum_{j=1}^n 
k_j,
\eeq
where $\{k_j =k(\lambda_j); j=1,\ldots,n\}$ are the roots obtained 
from the BAE 
\beq \label{H8}
e^{ik_jL} = - \prod_{l=1}^n \frac
{\sinh(\lambda_j-\lambda_l - i2\pi/3)}
{\sinh(\lambda_j-\lambda_l + i2\pi/3)}, \quad j=1,\ldots,n, 
\eeq
with
\beq \label{H9}
e^{ik_j} = \frac
{\sinh\lambda_j -i\sqrt{3t_p^2 +(4t_p^2-1)\sinh^2\lambda_j}}
{t_p(\sinh \lambda_j + i \sqrt{3}\cosh \lambda_j)}.
\eeq
As we see from \rf{H9} the left-hand side of the 
 BAE \rf{H8}  are quite distinct from the corresponding equations for other 
exact integrable quantum chains. In order to consider the bulk limit 
($L \to \infty$) it is then necessary to study these equations for 
small lattice sizes. These studies, as we are going to see in the next 
section,  will give us educated guesses for the 
topology of the roots $\{\lambda_j\}$ related to the low lying eigenvalues.   

 Before closing this section it is interesting to consider the BAE \rf{H8}-\rf{H9} at $t_p=1$. In this case they are given by 
\bea \label{H10}
&&\left[ \frac{\sinh(\lambda_j - i\pi/3)} {\sinh(\lambda_j+i\pi/3)}
\right]^L = - \prod_{l=1}^n \frac
{\sinh(\lambda_j - \lambda_l - i2\pi/3)}
{\sinh(\lambda_j - \lambda_l + i2\pi/3)}, \quad j=1,\ldots,n.
\eea
These equations coincide with the BAE of the XXZ chain \cite{yang} 
at the special value of its anisotropy $\Delta = -(q + 1/q) = 1/2$ 
($q=e^{i2\pi/3}$). However, in the XXZ chain the density of particles is restricted to 
$0 \leq \rho \leq 1$ while in the model \rf{H1} $0\leq\rho\leq2$. Although 
the completeness of the Bethe ansatz solutions is always a complicated 
problem, we expect that at $t_p=1$ the solution obtained from \rf{H10} is 
not complete.  Due to the additional symmetry (conservation of the number 
of pairs of particles) we should start again the Bethe ansatz \cite{AB1} or the matrix product 
ansatz \cite{MPA} taking into account this new symmetry. In this case we obtain 
the nested BAE of the deformed $SU(3)$ Perk-Schultz model with the 
deformation parameter value $q = e^{i2\pi/3}$. At this point this last 
model is special. In \cite{alcastro} several conjectures about the 
eigenspectra of this model on its antiferromagnetic regime was done. 
As is well known the eigenspectra of the anisotropic $SU(3)$ Perk-Schultz
 model  contains all the eigenvalues of the XXZ quantum  chain with the 
same anisotropy. The equivalence \rf{H5} then indicates that the whole 
eigenspectra of the XXZ with anisotropy $\Delta = 1/2$ is contained 
in the eigenspectra of the Hamiltonian \rf{H1} at $t_p=1$. 
 Moreover, a direct  diagonalization of \rf{H1}, with $t_p=1$ and $h=0$, 
shows that  for 
small even lattice sizes ($L\leq 12$)  the low lying 
eigenvectors, for densities $\rho <1$, coincide with those of the XXZ chain at the 
anisotropy $\Delta = 1/2$. This means that there is no double occupancy 
of particles on these eigenstates. This can be understood from \rf{H5} due to the 
presence of the magnetic field favoring single occupied sites. 

\section{Eigenspectra calculations}

In this paper we restrict ourselves to the cases where the 
parameter $t_p$ is real and positive. 
The eigenvalues of $H(t_p,h)$ are the same as those of $-H(-t_p,-h)$. 
The Hamiltonian \rf{H1}  although having a real trace is non hermitian. The exact 
eigenspectra calculations for lattice sizes $L \leq 12$ show 
that most of the eigenenergies are real.  All the low lying 
eigenvalues being real numbers. Imaginary eigenvalues, 
appearing in complex-conjugates pairs, happen only for 
high excited states in the eigenspectrum. 
 We also verify that when  $h=0$ 
the ground state of \rf{H1} belongs, for general values of 
$t_p$, to the sector with density $\rho =1$. Moreover, for 
$h=0$, the low lying excited states in the sectors with 
densities $\rho = 1 + \frac{m}{L}$ and 
$\rho = 1 - \frac{m}{L}$ ($m=1,2,\ldots$) are degenerated. 
This degeneracy is not valid for higher excited states 
since the model \rf{H1} at $h=0$ does not have the 
symmetry under the conjugation of particles: $0 \leftrightarrow 2$, $ 1 \leftrightarrow 1$. 

At $t_p =1$, where the model \rf{H1} recovers the deformed 
$SU_q(3)$ Perk-Schultz model at  $q = e^{i2\pi/3}$ 
(see \rf{H6}) the model is massive (non zero gap). Our 
numerical results indicate the same massive behaviour for all 
values of $t_p \geq 0$ as long as $h=0$. As  $h$ decreases 
(negative values) it reaches the critical field   $h_c(t_p)$. 
For 
 $h\leq h_c(t_p)$ 
the ground state changes continuously  
 its density of particles $\rho = \rho(h) <1$. 
We do expect, in the plane ($t_p,\rho$) a phase diagram 
with massless (critical) behaviour for $\rho <1$. In this 
critical regime the long-distance fluctuations should be 
ruled by a underlying conformal invariant field theory. The 
conformal central charge $c$ of the continuum theory can be 
estimated from the finite-size corrections of the ground-state energy $E_0(L,t_p,\rho)$ of the finite-size $L$ chain 
\cite{cardy-affleck}, i. e., 
\beq \label{H11}
\frac{E_0(L,t_p,\rho)}{L} = e_{\infty} - \frac{\pi v_s c}{L^2} + o(L^{-2}),
\eeq
where $e_{\infty}$ is the bulk limit of the ground-state 
energy per site and $v_s = v_s(t_p,\rho)$ is the sound 
velocity that can be inferred from the energy-momentum 
dispersion relations. Moreover,  for each primary operator 
$\Phi_{\Delta,\bar{\Delta}}$, with dimension 
 $x_{\Phi} = \Delta + \bar{\Delta}$ and spin $s_{\Phi} = 
\Delta -\bar{\Delta}$ in the operator algebra of the underlying conformal theory, there exists an infinite tower of 
eigenstates, whose energies $E_{m,m'}^{\Phi}(L)$ and 
momentum $P_{m,m'}^{\Phi}$ behave asymptotically 
as \cite{cardy} 

\bea \label{H12}
&&E_{m,m'}^{\Phi}(L) = E_0(L) + \frac{2\pi}{L}v_s 
(x_{\Phi} +m +m') + o(L^{-1}), \nonumber \\
 && P_{m,m'}^{\Phi } = \frac{2 \pi }{L} 
(s_{\Phi} +m -m'), 
\eea
with $m,m' = 0,1,2,\ldots$.

Our spectral analysis of the Hamiltonian \rf{H1} was done 
by solving numerically the BAE \rf{H8}-\rf{H9} using a 
Newton type method, whenever it was possible. Since there is 
no numerical method that warranties the  solution for the non linear 
equation \rf{H8}, the success in finding the 
solutions depends very much on the ability to provide 
educated guesses for the approximated values.  In the 
cases where we were not able to obtain the solutions of the BAE 
our 
analysis were based on direct numerical diagonalizations, 
or by using approximated methods like the power method. On 
these last cases our analysis were limited for lattice sizes up to 
$L=24$ sites. 

\begin{figure}[ht!]
\centering
{\includegraphics[angle=0,scale=0.46]{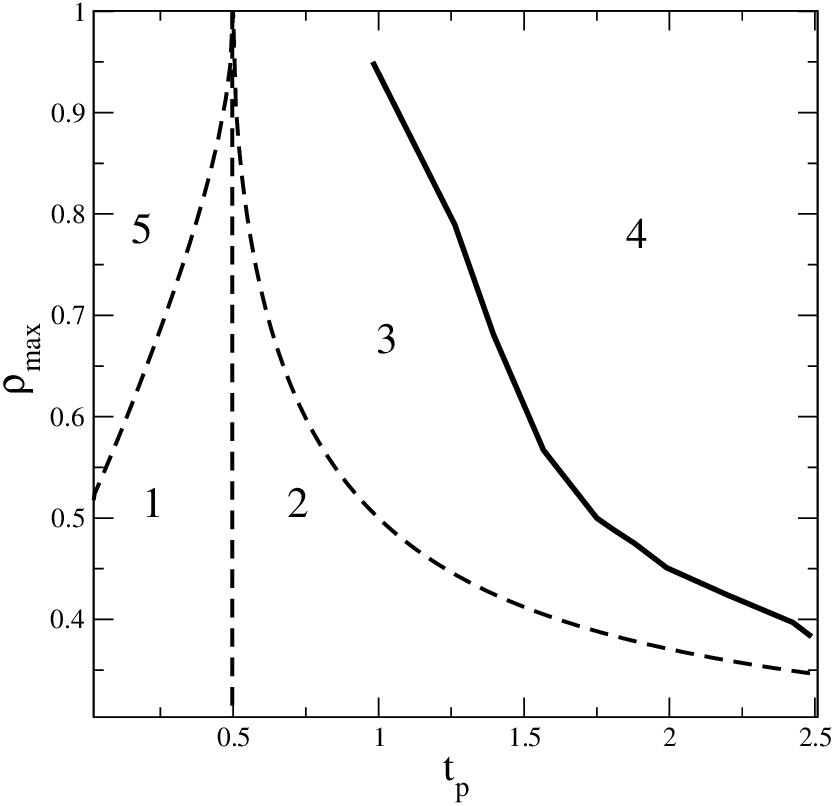}}
\caption{Phase diagram of the quantum Hamiltonian \rf{H1}-\rf{H3}. In 
regions 1 and 2 the ground state is described by real roots of the BAE. 
In the other regions the ground state, besides real roots, also contains 
complex ones. Regions 1,2,3 and 5 are critical and governed by $c=1$ 
Coulomb gas conformal field theory. Region 4 is characterized by 
several crossing of the eigenenergies producing  an oscillatory behaviour 
in the finite  energy gaps and momentum.  
The lines separating regions 1 and 5, and regions 2 and 3 are obtained 
by solving \rf{H18}-\rf{H19}  and \rf{H21}. An schematic line (heavy line)  where we expect 
 a first-order 
phase transition, separating region 3 and 4,   is 
 also shown.}
\label{fig1}
\end{figure}

According to the different topologies of roots of the BAE 
\rf{H8}-\rf{H9} we divide the plane ($t_p,\rho$) in five 
regions (see Fig. \ref{fig1}).
%

In general the roots $\{\lambda_j\}$ of \rf{H8} are 
complex numbers. Numerical analysis, based on direct 
solutions of \rf{H8},  as compared with brute force 
diagonalizations of the quantum chains shows that in regions 
1 and 2 (see Fig~\ref{fig1}) the roots $\{\lambda_j\}$ 
corresponding to the low lying eigenvalues are all real 
numbers. For these cases,  the right-hand side of 
\rf{H8}  is unimodular. The roots are then  
constrained or not on an 
 finite real interval, depending on the values of $t_p$. If $t_p <1/2$ 
(region 1) the real roots are constrained on the interval 
\beq \label{H13p}
-\Omega(t_p) <\lambda_j < \Omega(t_p),
\eeq
where 
\beq \label{H13}
\Omega(t_p) = \frac{1}{2} \cosh^{-1}(\frac{2t_p^2+1}{1-4t_p^2} ).
\eeq
For $t_p \geq 1/2$ (region 2) these roots are 
unconstrained. For these real roots the BAE \rf{H8}-\rf{H9} 
take the simple from 
\beq \label{H14}
Lk(\lambda_j) = 2\pi Q_j^{(n)} + \sum_{k=1}^n \phi(\lambda_j,
- \lambda_k), \quad j=1,\ldots,n, 
\eeq
where 
\bea \label{H15}
&& k(\lambda) = \theta_2(\lambda) - \theta_1(\lambda), \quad 
\phi(\lambda) = 2 \arctan (\frac{\tanh \lambda}{\sqrt{3}}), 
\nonumber \\
&& \theta_2(\lambda) = - \arctan (\frac{\sqrt{3t_p^2+ (4t_p^2-1)\sinh^2\lambda 
}}{\sinh{\lambda}}), \nonumber \\
&& \theta_1(\lambda) = \arctan (\sqrt{3}\coth \lambda),
\eea
and $Q_j^{(n)}$ ($j=1,\ldots,n$) are integers or odd-half 
integers, depending on the particular eigenstate. As before 
the eigenenergies and momenta  are given by \rf{H7}.
 
In order to illustrate we present in table~\ref{table1} some eigenenergies  obtained 
by solving the BAE \rf{H14} in region 1. We take in \rf{H1} the 
magnetic field $h=0$.  
 They are obtained for several lattice sizes $L$ at $t_p=0.2$, $0.3$ and $0.4$ and densities $\rho=0.6$, 
$0.65$, and $0.7$, respectively. 

\begin{figure}[ht!]
\centering
{\includegraphics[angle=0,scale=0.46]{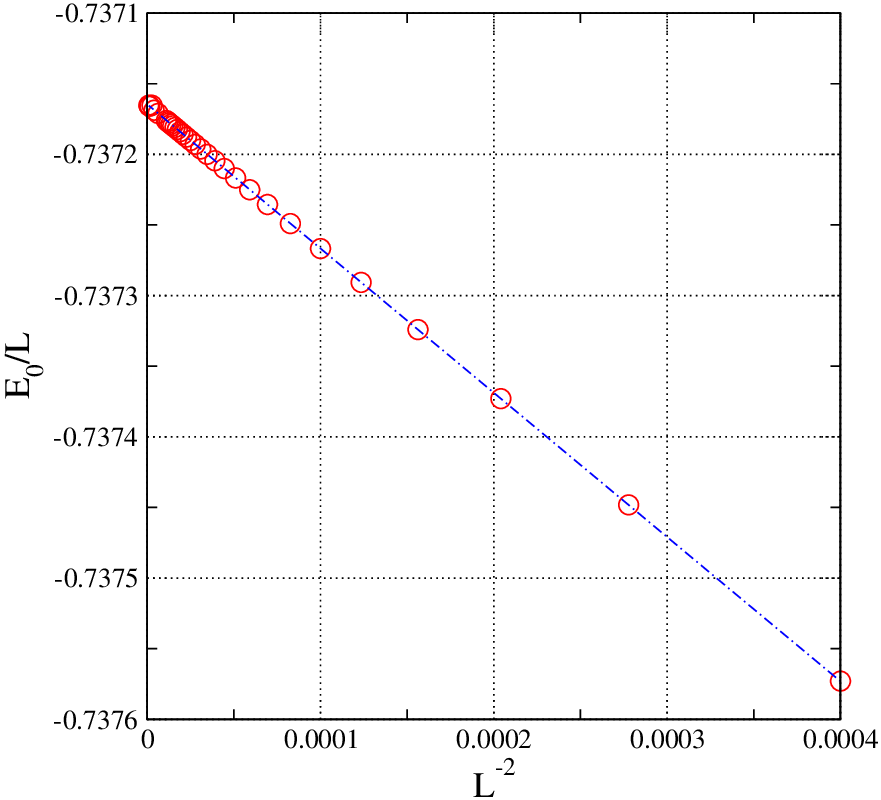}}
\caption{ Ground state energy per site of the Hamiltonian \rf{H1} as a 
function of $1/L^2$, for $t_p =0.2$,  $\rho=0.6$. We set $h=0$ in the 
figure. This point belongs to region 1 in Fig.~\ref{fig1}.}
\label{fig2}
\end{figure}

\begin{table}[hbt]
\centering
\begin{tabular}{|c||c|c|c|}
\hline
$L \setminus (t_p,\rho)$ & (0.2,0.6)  & (0.3,0.65) & (0.4,0.7) \\ \hline \hline
    10  &\,-2.24745586\, & \, -\, & \,-1.70579604\,\\ \hline
    100 &\,-2.23726600\, & \, -1.86555023\, & \,-1.69990939\,\\ \hline
    200 &\,-2.23719021\, & \, -1.86548482\, & \,-1.69653825\,\\ \hline
    1000 &\,-2.23716569\, & \, -1.86547050\, & \,-1.69651539\,\\ \hline
    Ext. &\,-2.23716439\, & \, -1.86546561\, & \,-1.69624038\,\\ \hline
\end{tabular}
\caption[table1]{Lowest eigenenergies per site 
  of the Hamiltonian \rf{H1} 
with $h=0$ for 
some values of $L$, $t_p$ and density of particles $\rho$. 
These points belong to region 1 (see Fig.~\ref{fig1}). They 
are obtained by solving directly the BAE \rf{H14}. They last line is 
the asymptotic value obtained from the solution of \rf{H18}-\rf{H20}.} 
\label{table1}
\end{table} 
 They are zero (mod. $\pi$) momentum eigenstates with roots 
$\{\lambda_j\}$ symmetrically distributed around the origin. 
The corresponding quantum  numbers in \rf{H15} are $Q_j^{(n)} = \pm (
-\frac{L}{2} +j)$ ($j= 1,\ldots,n=\rho L$). 
 The finite-size corrections of the ground state energies indicate us 
that in region 1 and 2 the Hamiltonian \rf{H1} is critical 
and conformally  invariant. The relation \rf{H11} gives us an 
estimate for the  
conformal anomaly $c$ of the underlying conformal field 
theory. In Fig.~\ref{fig2} we show, as an example, the 
ground-state energy per site as a function of $1/L^2$, for 
the quantum chain with $t_p = 0.2$,  density $\rho = 0.6$ and $h=0$. 
We clearly see a linear behaviour as predicted by conformal 
invariance (see \rf{H11}). The linear coefficient of the dot 
line in Fig.~\ref{fig2} gives us, from \rf{H11}, an estimate 
for the product $2\pi v_sc$. The sound velocity $v_s$ is more 
difficult to estimate from numerical solutions of 
\rf{H14}-\rf{H15}, since it demands the calculation of 
excited states with non zero momenta. The BAE roots on this 
case are not symmetric. A possible way to circumvent this 
problem is to seek for excited zero momentum states related 
to higher values of $m=m'$ in the conformal tower \rf{H12} 
of the identity operator ($x_{\phi}=0$). 
Our numerical analysis indicate that these energies have a 
configuration of BAE roots where $\rho L-2$ of them are real 
and symmetrically distributed around the origin,  and the two out-most roots  are in the form $\lambda = \pm(a + i\pi)$, with 
$a \in \Re$. 
There is a  difficulty in using this set of eigenlevels. As   we 
change  the lattice size their  relative positions $m=m'$ in the conformal 
tower are not fixed. However the estimates for the sound velocity 
obtained by direct diagonalization ($L \leq 24$) 
 of the quantum chain, although with 
low precision, are enough to indicate the position $m=m'$ of  the 
eigenlevel in the conformal tower. Using this procedure we computed 
the conformal anomaly at several points in regions 1 and 2 
(see Fig.~\ref{fig1})  obtaining $c=1.00(1)$. 

Since in regions 1 and 2 the ground state is described  by real roots 
 of the BAE, 
it is not difficult, in this case, 
 to consider the bulk limit ($L\to \infty$) of 
the BAE \rf{H14}-\rf{H15}. Following Baxter \cite{baxter} we 
define the variables
\bea \label{H16}
&&x_j = \frac{Q_j}{L} = \frac{1}{2\pi}\left[
k(\lambda_j) - \frac{1}{L}\sum_{l=1}^{\rho L} \theta(\lambda_j-\lambda_l)
\right].
\eea
When $L\to \infty$, $x_j \to x$ become a continuous variable in the 
interval $x_{\mbox{\scriptsize min}} < x < x_{\mbox{\scriptsize max}}$ 
satisfying the integral equation
\bea \label{H17}
&& x_j = \frac{Q_j}{L} = \frac{1}{2\pi}\left[k(\lambda_j) - 
\int_{x_{\mbox{\scriptsize min}}}^{x_{\mbox{\scriptsize max}}} 
\phi(\lambda(x) -\lambda(y))dy\right].
\eea
Since for the ground state the $\{Q_j\}$ are equally spaced by the 
unity, $\sigma_L(\lambda) = \frac{dx}{d\lambda}$ will give us, 
for $L \to \infty$, the local density of roots $\sigma_{\infty}
(\lambda)$, that satisfies the integral equation
\bea \label{H18}
&&\sigma_{\infty}(\lambda) = \frac{1}{2\pi}\left( \frac{dk(\lambda)}{d{\lambda}} + 
\int_{-\lambda_0}^{\lambda_0} \frac{2\sqrt{3} \sigma_{\infty}(\lambda^\prime)}{2\cosh (2(\lambda-\lambda^\prime)) 
+1}  d\lambda^\prime \right),
\eea
where from \rf{H15}
\bea \label{H18p}
&&\frac{dk(\lambda)}{d\lambda} = \frac{\sqrt{3}}{2\cosh (2\lambda) +1}
\left(1 + \frac{\sqrt{3}\cosh \lambda}{\sqrt{(4t_p^2-1)\sinh^2\lambda + 
3t_p^2}}\right),
\eea
and $\lambda_0 = \lambda_0(t_p,\rho)$ gives the extreme values of the 
roots. The total density of particles and the ground-state energy per 
site are given by 
\beq \label{H19}
\rho = \int_{-\lambda_0}^{\lambda_0}  \sigma_{\infty}
(\lambda) d\lambda,
\eeq
\beq \label{H20}
\epsilon_0(t_p,\rho,h) = \epsilon_0(t_p,\rho,0) - h\rho = 
-\rho ( \frac{1}{2t_p} +h) + 
2\int_{-\lambda_0}^{\lambda_0} \cos((k(\lambda))\sigma_{\infty}
(\lambda) d\lambda,
\eeq
where $h=h(t_p,\rho)$ 
is the magnetic field that fixes the ground-state energy at the 
density $\rho$. 

\begin{figure}[ht!]
\centering
{\includegraphics[angle=0,scale=0.46]{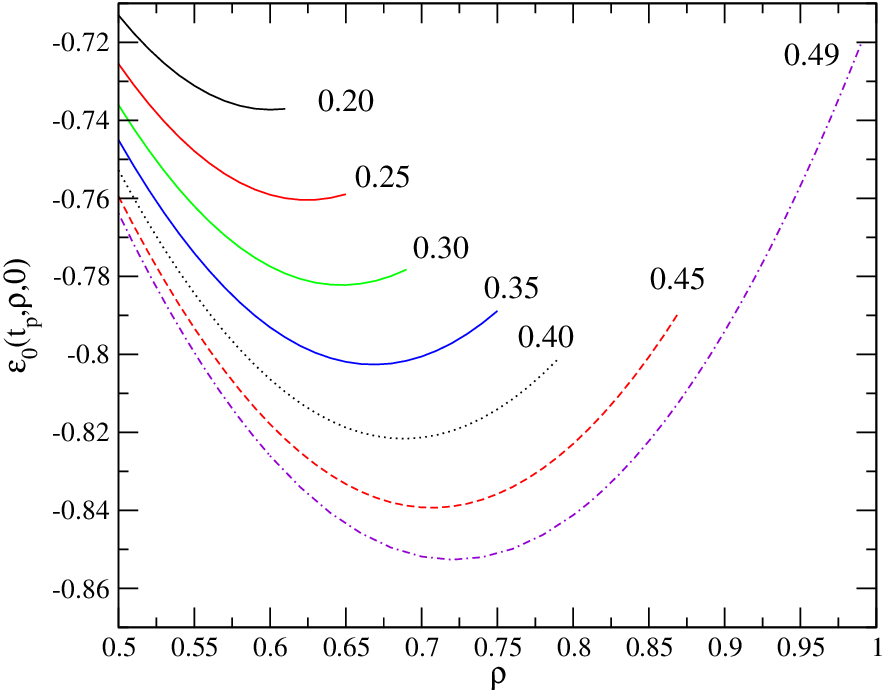}}
\caption{Ground state energy per site of the Hamiltonian \rf{H1} as a 
function of the density $\rho$, for several values of $t_p$ in 
region 1. 
In the figure we set $h=0$ and the
 values 
of $t_p$ are shown. These curves are obtained by solving \rf{H18}-\rf{H19}. For a given value of $t_p$ the curves are shown for 
$\rho <\rho_{\mbox{\scriptsize max}}(t_p)$, with $\rho_{\mbox{\scriptsize max}}(t_p)$  given by \rf{H21}. 
For $\rho >\rho_{\mbox{\scriptsize max}}(t_p)$ there exist 
complex roots in the BAE solutions for the ground state and the 
model is in region 5.}
\label{fig3}
\end{figure}

For $t_p <1/2$ (region 1), $\lambda_0(t_p,\rho)$ is 
always finite, i. e., $\lambda_0(t_p,\rho,h) \leq \Omega(t_p)$ 
 (see \rf{H13}). 
 In Fig.~\ref{fig3} we show the ground-state energy 
$\epsilon_0(t_p,\rho,h)$ 
for several values of $t_p$ in region 1 (we set $h=0$ in the figure). 
They are obtained by using in 
\rf{H20} the $\sigma_{\infty}(\lambda)$ obtained from the numerical 
solution of the coupled integral equations \rf{H18}-\rf{H19}. As 
we see from this figure these equations give us a maximum total density 
of particles $\rho = \rho_{\mbox{\scriptsize max}} (t_p) <1$ for the 
quantum chain. At the endpoints of the curves 
we have  $\lambda_0 = \Omega(t_p)$ 
(see \rf{H13}). Above this density, which we refer as region 5, some 
of the 
BAE roots associated to the ground state of the quantum chain 
have complex values. In Fig.~\ref{fig1} the line separating
 regions 1 and 5  gives, for a given $t_p$, the 
maximum density obtained from \rf{H18}-\rf{H19}. 
 This line was obtained by solving numerically 
 \rf{H18}-\rf{H19}.
In order to compare 
with the finite-size results we give in the last line of table \ref{table1} 
the estimated value of $\epsilon_0(t_p,\rho,h=0)$, obtained by solving 
\rf{H18}-\rf{H19} for $t_p =0.2, \; 0.3$ and $0.4$ with 
$\rho =0.6,\; 0.65$ and $0.7$, respectively.

For $t_p\geq 1/2$ (region 2) the roots $\{\lambda_j\}$ are not 
constrained and $\lambda_0=\lambda_0(t_p,n)$ is arbitrary. The maximum 
density compatible with only real roots of the BAE for 
the ground state is obtained by setting $\lambda_0 \to \infty$ in 
\rf{H18} and \rf{H19}. In this case we can solve \rf{H18}-\rf{H19} by 
using Fourier transforms. After some long, but straightforward 
calculation we obtain the maximum density
\bea \label{H21} 
&&\rho_{\mbox{\scriptsize max}}(t_p) = \int_{0}^{\infty} 
\frac{3\sqrt{3}}{2\cosh(2\lambda)+1} \left(1 + \frac{\sqrt{3}\lambda}
{\sqrt{(4t_p^2-1)\sinh^2\lambda +3t_p^2}}\right)d\lambda.
\eea
At special values of $t_p$ we are able to solve \rf{H21} analytically:
$\rho_{\mbox{\scp max}} (1/2) = 1$, $\rho_{\mbox{\scp max}} (1) = 1/2$, $\rho_{\mbox{\scp max}} (\infty) = 1/4$. In Fig.~\ref{fig1} the curve separating  region  2 from  3   was obtained from the numerical 
evaluation of \rf{H21}. As happened in region 5, for 
$\rho >\rho_{\mbox{\scp max}}(t_p)$ (regions 3 and 4), the BAE solutions 
giving the ground state contain complex roots besides the  real ones. 

As in region 1 (see table \ref{table1}) 
 we also calculated the finite-size corrections for the 
ground-state energy in several points of region 2. Using \rf{H11} and the 
same procedure as before our results indicate that, like  region 1, 
 region 2 is also critical and conformal invariant with $c=1$.   
  We then have in both regions (1 and 2) a massless behaviour 
 with ground-state energy given by real roots of the BAE. This 
 critical behaviour  is quite similar as that of the XXZ quantum chain in the 
presence of a magnetic field \cite{woyna,xxz-h}. We then expect in 
regions 1 and 2 a physical behaviour described by an underlying $c=1$ Coulomb gas conformal field theory.
 The anomalous dimensions of operators being given by 
\beq \label{H22} 
x_{l,m} = l^2x_p + m^2/4x_p,
\eeq
with $l,m=0,\pm1,\pm2,\ldots$. The dimensions $l^2x_p$ are obtained from
 \rf{H12}, by considering the difference between the ground-state 
energies in the sectors with densities  $\rho$ and $\rho + l/L$. The 
dimensions $x_{0,m}$ are calculated by using in \rf{H12} the mass gaps 
associated to eigenstates with the same  density of particles. 

\begin{figure}[ht!]
\centering
{\includegraphics[angle=0,scale=0.46]{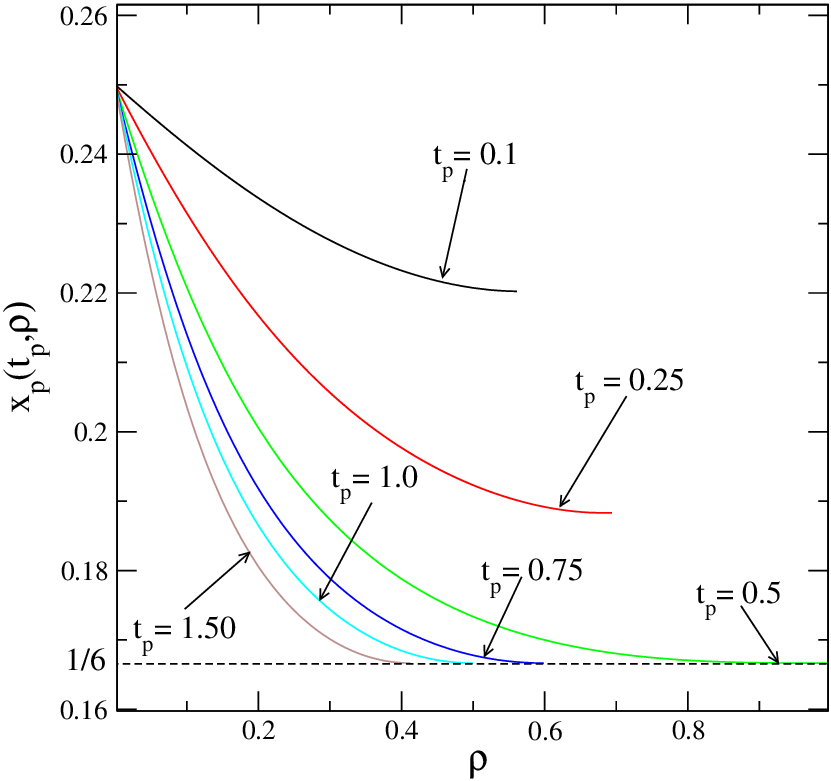}}
\caption{ The values of $x_p(t_p,\rho)$ (see \rf{H22}) as a function of $\rho$ for several values of $t_p$ in regions 1 and 2. 
The curves with $t_p <1/2$ and $t_p \geq 1/2$ belong to regions 1 
and 2 of Fig.~\ref{fig1}, respectively. The endpoints
 of the curves are the densities separating regions 1 and 2 from 3 and 5, 
respectively. For $t_p \geq 1/2$ the endpoints of the curves is $x_p= 1/6 \sim  0.16666$.}
\label{fig4}
\end{figure}

\begin{figure}[ht!]
\centering
{\includegraphics[angle=0,scale=0.46]{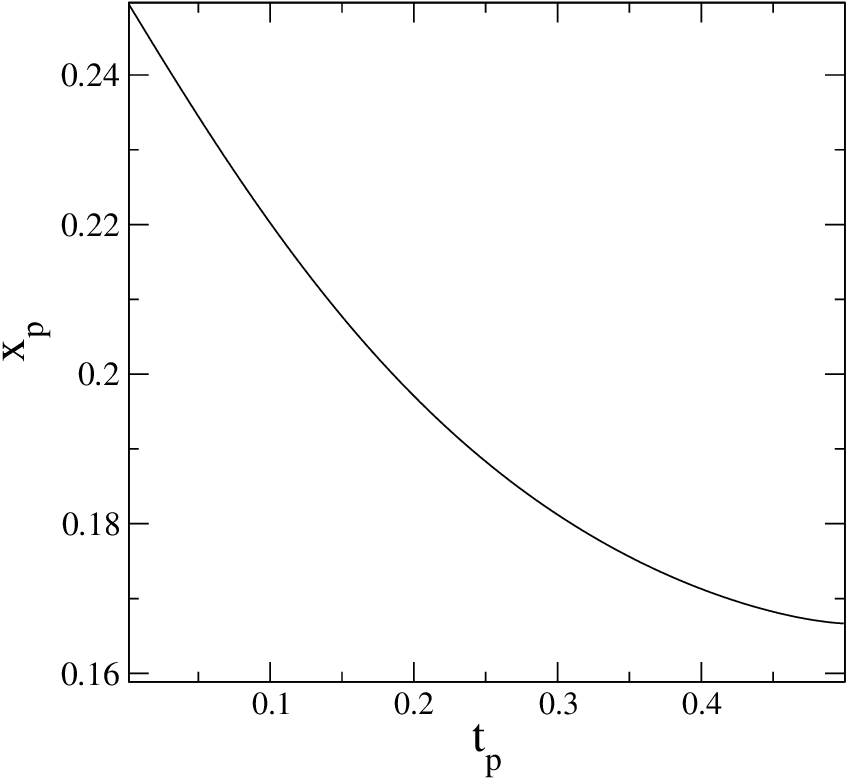}}
\caption{ 
The values of $x_p$ at the line separating regions 1 and 5 
(see Fig.~\ref{fig1}). The exponents varies from $x_p=1/4$ to $x_p=1/6\sim 0.1666$, as $t_p$ goes from 
0 to 1/2.}
\label{fignew}
\end{figure}

The parameter $x_p=x_p(t_p,\rho)$ in regions 1 and 2, which are  
estimated from the finite-size corrections of the energy \rf{H20}, 
can be calculated analytically. This is done by applying to our 
relations \rf{H17}-\rf{H20} the method used in \cite{woyna} for the 
XXZ quantum chain in a magnetic field. We obtain
\beq \label{H23}
x_p = (2\xi(\lambda_0))^{-2},
\eeq
where $\xi(\lambda_0)$ is the dressed charge \cite{korepin} evaluated 
at the Fermi surface $\lambda_0$ of the effective Coulomb gas. This 
function  satisfy the integral equation
\beq \label{H24}
\xi(\lambda) = 1 + \frac{\sqrt{3}}{\pi} \int_{-\lambda_0}^{\lambda_0} 
\frac{\xi(\lambda^\prime)}{2\cosh(\lambda - \lambda^\prime)+1} d\lambda^\prime.
\eeq
In Fig.~\ref{fig4}  we show, for several values of $t_p$ in 
regions 1 and 2, the dimensions $x_p = x_p(t_p,\rho)$ obtained by 
using in \rf{H23} the numerical solutions of the coupled integral 
equations \rf{H18}, \rf{H19} and \rf{H24}.
We can see from this figure that for any $t_p$, as $\rho \to 0$, 
$x_p \to 1/4$. 
This can be understood from the fact that at this limit the 
interacting potential energy is negligible when compared with the 
kinetic energy (hopping terms). We have essentially non-interacting 
particles, where $x_p$ has the value  1/4.
 We also see from this figure that, for $t_p<1/2$ (region 1), the limiting 
value of $x_p$ depends on the value of $t_p$. This value is 
obtained by choosing $\lambda_0=\Omega(t_p)$ in \rf{H24}, where 
$\Omega(t_p)$ is given by \rf{H13}. In Fig.~ \ref{fignew} we show 
the limiting values of $x_p$ as a function of $t_p$, for 
$t_p \leq 1/2$. 
%
%
%
It varies from $x_p=1/4$ to $x_p=1/6$ as $t_p$ goes 
from 0 to 1/2. Moreover, Fig.~\ref{fig4}  shows that the limiting value 
is $x_p=1/6$ for any $t_p \geq 1/2$. This should be the case since for 
any $t_p\geq 1/2$ the maximum value of $\lambda_0$ is infinity and 
consequently 
\rf{H23} and \rf{H24}  give us  the same result for any value of $t_p$. 
  The exact value $x_p=1/6$ can be understood from the 
relation (see Sec. II) 
of the model at $t_p=1$ and the XXZ chain with anisotropy 
$\Delta=1/2$. When $\lambda_0 \to \infty$ the density $\rho \to 
1/2$. At this density the XXZ has  no magnetic field and its exponent 
is given by $x_p=(\pi-\cos^{-1} (-\Delta))/2\pi =1/6$.  
These results imply that, at
 the line 
separating regions 1 and 5,  $x_p$ varies continuously from 1/4 to 
1/6 and, in the line separating regions 2 and 3, it remains  fixed to the 
value 1/6.

In regions 3 and 5 of Fig.~\ref{fig1} some of the roots of the BAE 
corresponding to the ground state have complex  values. In fact in some 
points of these regions we were able, for small lattice sizes, to solve 
the BAE for the ground-state energy. We verified that besides real 
roots we also have pairs of 2-strings (pair of roots of type 
$\lambda_{\pm} = a +ib$, with $a,b \in \Re$). The mixture of real and 
complex roots in the BAE \rf{H8} produces numerical instabilities 
turning the numerical solution of the BAE a quite difficult task. Due to 
this difficulty, instead of solving the BAE \rf{H8} we have used 
the power method to calculate the lower part of the eigenspectrum.  On 
this case, due to computer limitations, even exploring the $U(1)$ and 
translation symmetries of the quantum chain we are restricted to lattice 
sizes up to $L=24$, where the largest sector we can calculate  the 
lowest eigenenergy has dimension 5,136,935 ($L=24$, $n=13$). 
Since the calculations are done for a fixed density 
of particles $\rho$ the larger number of sites we can use, and 
consequently get a better precision for our estimates, is at the density 
$\rho = 1/2$. 
\begin{table}[hbt]
\centering
\begin{tabular}{|c||c|c|c|c|c|c|c|c|}
\hline
$t_p$ &  $e_{\infty}$ & $v_s$  & $h$ & c & $x_{+1,0}$ & $x_{-1,0}$ & $x_{0,1}$ & $1/4x_{0,1}$ 
         \\ \hline \hline
 \,0.75\, &  \,-1.11947\, & \,1.4321\,& \,-1.5911\, &\,1.000\, & \,0.168\, & \,0.169\, & \,1.490\, & \,0.168\,     
 \\ \hline 
 1.0 &   -1.04904  & 1.2991 & -1.5000 & 0.999 & 0.167 & 0.167 & 1.500 &
0.167     
 \\ \hline
 1.15 &   -1.02211  &  1.2252 &-1.4686 &  0.999 & 0.170 & 0.169 & 1.498 &
0.167     
\\ \hline
 1.25 &   -1.00792  &  1.1771 &-1.4540 &  1.001 & 0.169 & 0.167 & 1.490 &
0.168     
\\ \hline
 1.35 &   -0.99597  &  1.1300 &-1.4425 &  1.001 & 0.169 & 0.167 & 1.490 &
0.168     
\\ \hline
 1.50 &   -0.9812  &  1.0596 &-1.4295 &  1.01 & 0.169 & 0.167 & 1.48 &
0.168     
\\ \hline
\end{tabular}
\caption[table1]
{The quantities $e_{\infty},v_s,c,x_{+1,0},x_{-1,0}$ and $x_{0,1}$ 
for the Hamiltonian \rf{H1} with density $\rho=1/2$ and some values of 
$t_p$. The first line ($t_p=0.75$) corresponds to a point inside region 2 
and the second one ($t_p=1$) to a point at the line separating regions 2 
and 3. The remaining lines correspond to points in region 3 (see Fig.~\ref{fig1}).}
\label{table2}
\end{table}

In table~\ref{table2} we present some of our  estimated values of   
several quantities that characterizes the critical behaviour of 
 the quantum chain \rf{H1} inside region 3 
($t_p=1.15$, $1.15,1.35$ and $1.5$). The calculations were 
done at density $\rho =1/2$. 
In the first two lines, for comparison, we   
also give the results for  a point inside region 2 ($t_p=0.75$) 
 and a point at the line separating 
regions 2 and 3 ($t_p=1$). 
We include in the table the estimated values for 
the ground-state energy per particle in the bulk 
limit $e_{\infty}$ (we set $h=0$ in \rf{H1}),
 the sound velocity $v_s$ and the 
magnetic field $h$ that fixes the ground state at density $\rho=1/2$.

The value of the exponent $x_{+1,0} = 
x_{-1,0}=x_p$ for $t_p=0.75$ and $t_p=1$ are known from the 
solutions of \rf{H18}, \rf{H19} and \rf{H24}, i. e., $x_p= 0.16770$ 
($t_p=0.75$) and 
$x_p=0.1666$ ($t_p=1$). The comparison of these last results with the 
first two lines of table~\ref{table2} indicates that the errors are 
in the last digit. 
Like regions 1 and 2, region 3 is 
also massless with a conformal central charge $c=1$. The dimensions $x_{+1,0}$ and 
$x_{-1,0}$ in table~\ref{table2} are associated to the sectors with 
densities $\rho =\frac{1}{2}+\frac{1}{L}$ and $\rho = \frac{1}{2}-
\frac{1}{L}$, 
respectively. These dimensions in a standard Coulomb gas phase, like in 
region 1 and 2, are equal and correspond to the dimensions $x_{\pm1,0}$ in \rf{H22}.
We also show in  table~\ref{table2} the dimension 
 $x_{0,1}$ related to the first gap with zero (mod. $\pi$) momentum in the 
sector containing the ground state. The dimensions $x_{\pm1,0}$, 
as shown in the table, is 
 close to $1/4x_{0,1}$, in agreement with \rf{H22}. 
The dimensions in the first two lines of table II  ($t_p=0.75$ and $t_p=1$) although close are not equal, as we can also check in Fig~\ref{fig4}.  Table II also indicate that inside region 3 the conformal dimensions almost does not change as we change $t_p$. Since our results are valid only for the 
density $\rho=1/2$ we do not know if this small (or no) variation 
remains  valid for other points inside  region 3.

\begin{figure}[ht!]
\centering
{\includegraphics[angle=0,scale=0.46]
{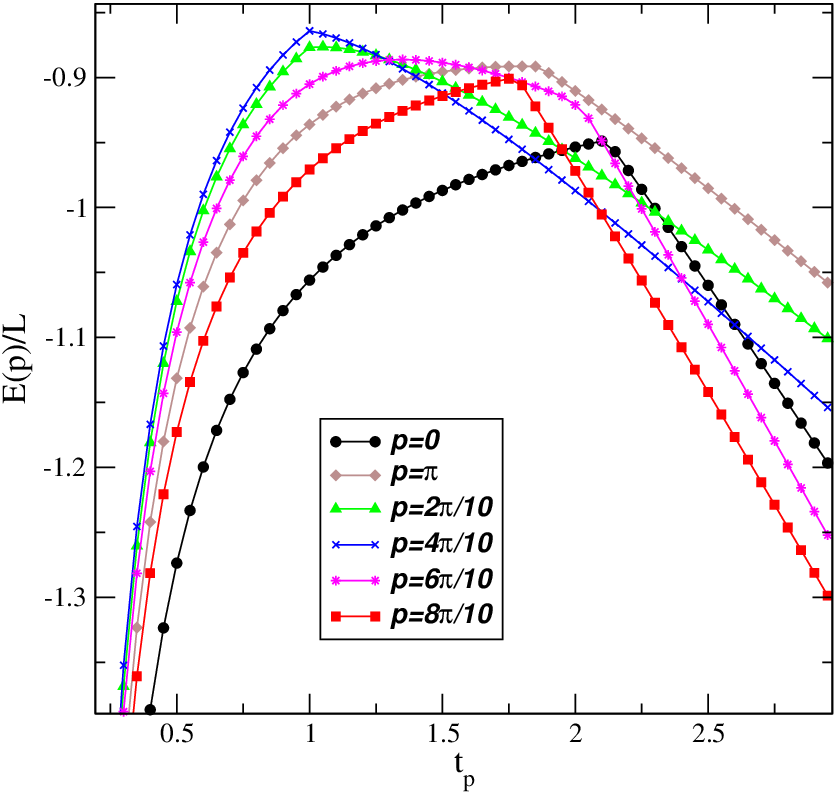}\hspace{-3.0cm}\hfill\includegraphics
[angle=0,scale=0.46] {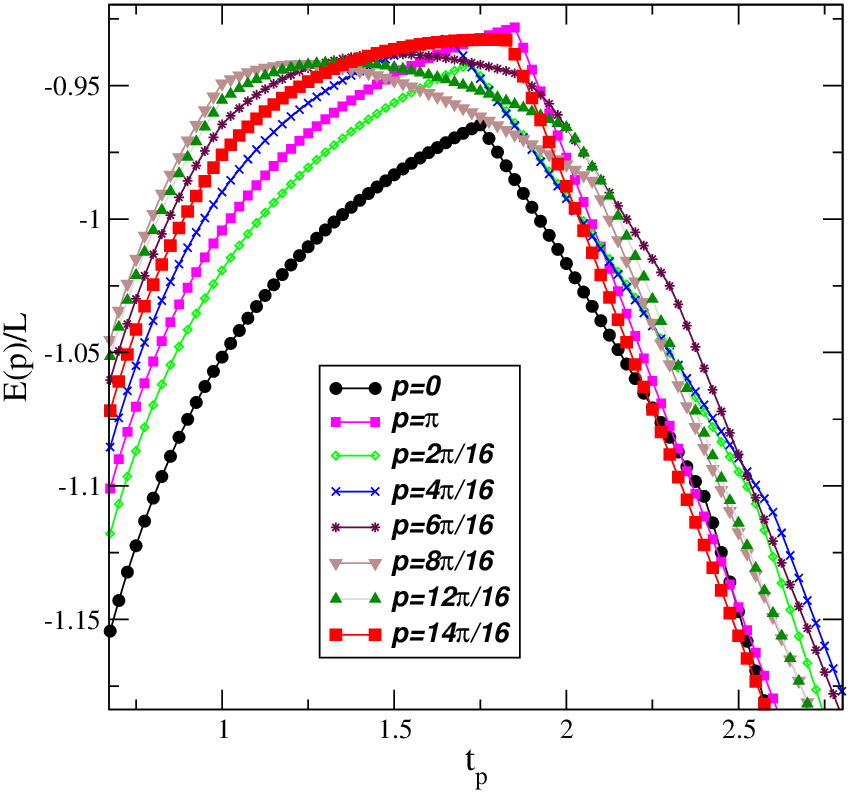} } 
\caption{Lowest eigenenergy $E(p)$ with momentum $p$, as a 
function of  $t_p$ for 
the Hamiltonian \rf{H1} at density $\rho=1/2$ and  lattice sizes $L=10$ 
(left) 
and $L=16$ (right). We set $h=0$ in the figure. The energies $E(p)$ 
are the lowest eigenenergies with momentum $p$. The eigenlevels with 
$p \neq 0,\pi $ are doubled degenerated (momentum $p$ and $-p$). The kinks on 
the curves are due to a level crossing among the two lowest eigenenergies 
with the same momentum.} 
\label{fig6}
\end{figure}

Once in region 3 of Fig.~\ref{fig1} by increasing the value of $t_p$, for 
a fixed density, we reach region 4. In this region several eigenlevels 
crossings happen in the finite lattice. 
In Fig.~\ref{fig6} we show for $L=10$ (left) and $L=16$ (right) these 
crossings. The crossings among eigenlevels with distinct momenta are 
visualized directly in the figure. The kinks in the curves are due to 
the level crossings with the same momentum. 

As we  cross from region 3 to 
region 4 the ground-state energy shows a discontinuity on its derivative 
due to a change of its relative position with an excited eigenstate. 
 Unfortunately also in this region, due to numerical instabilities, it is 
quite difficult to solve directly the BAE \rf{H8}. The direct 
diagonalization of the quantum chain, except at the density $\rho =1/2$, 
can only be done for quite few lattice sizes. Our estimate of the line 
separating regions 3 and 4, shown in Fig.~\ref{fig1}, is then just 
schematic. 
This indicates 
 a first order phase transition along the line separating these 
regions.

%

\begin{figure}[ht!]
\centering
{\includegraphics[angle=0,scale=0.56]{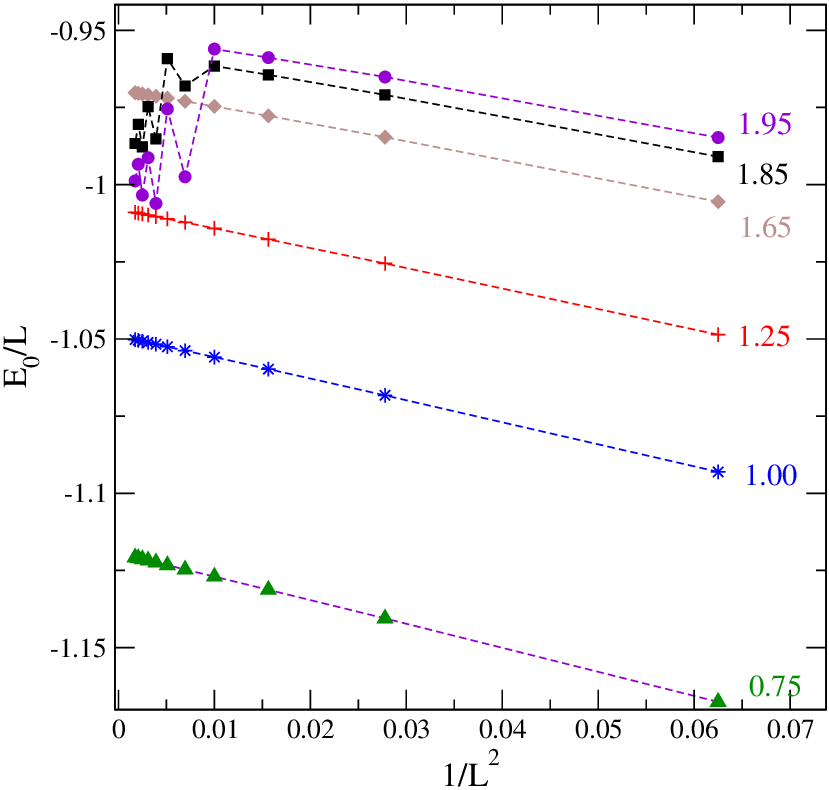}}
\caption{ Ground state energy per particle as a function of $L^{-2}$ 
for the Hamiltonian \rf{H1} with $h=0$ at density $\rho=1/2$. 
The lattice sizes are $L=6,8,\ldots,24$. The curves correspond to 
$t_p=0.25,0.5$ (region 1),  
$t_p=0.75,1.00$ (region 2), $t_p=1.25$ (region 3) and $tp=1.85,1.95$ 
(region 4).}
\label{fig7}
\end{figure}

Inside region 4 and for $\rho =1/2$ we verify that the leading finite-size 
correction of the ground-state energy is not $O(1/L^2)$ as expected  
in  a massless conformaly invariant phase (see \rf{H11}). This is illustrated in 
Fig.~\ref{fig7} where we show for some values of $t_p$ the ground-state 
energy as a function  of $1/L^2$ for the density $\rho=1/2$.  
We see 
in this figure the distinct behaviour for the points belonging to 
regions 2 and 3 as compared with those of region 4 (lower two curves).  
The oscillatory behaviour in Fig.~\ref{fig7} for the points inside 
region 4 is a consequence of the level crossings happening inside this region. As we can see in Fig.~\ref{fig6} tor $t_p \sim 1.65$ many crossings 
happen with the crossing position  dependent of the lattice size. 
As we change $t_p$ inside this region, for a given finite lattice $L$, 
the momentum of the ground state also changes. For example for 
$t_p \approx 2.25$ (see fig.~\ref{fig6}) the ground state 
for the quantum chains with 
$L=10$ and $L=16$ has a non zero momentum and is doubled degenerated.  The oscillatory behaviour 
shown in Fig.~\ref{fig7} turns out the finite-size scaling analysis 
quite imprecise. Although not conclusive these oscillatory behaviour in energy and momentum indicate that we have in region 4 effects of 
incommensurability of the density distributions,  similar as happening in other models \cite{income}.

In region 5, as in region 3, it is difficult to solve numerically 
the BAE due to the occurrence of complex roots  mixed with real ones. 
In this case our analysis was restricted to small lattice sizes 
($L\leq 24$). Since we should consider sequence of lattices with 
fixed density and on this region $\rho>1/2$, the number of data  
we can have on 
a given finite-size sequence is quite small. However, for a given lattice size 
and density, as we change $t_p$ we did not see the crossings of eigenlevels observed in region 4 
(see figure 
\ref{fig6}). This indicates that region 5, similarly as regions 1,2 and 3 
is a critical $c=1$ Coulomb gas phase.  
 We greatly welcome more 
 precise and convincing results for regions 4 and 5. 

\section{Summary and conclusions}

We have made a detailed analysis of the spectral properties and phase 
diagram of the one of the new spin one model introduced in \cite{AB1}. 
This model is 
exact integrable for arbitrary values of the parameter $t_p$. Our analysis 
was restricted to the cases where $t_p >0$. At $t_p =1$ we have shown that 
the model recovers the deformed $SU(3)$ Perk-Schultz model 
(or spin-1 Sutherland model)  at the special 
value of its deformation parameter $q=e^{i2\pi/3}$ and external magnetic 
field. We verified that at 
this point the low lying eigenvalues of the model are the same as the 
corresponding ones of the XXZ quantum chain with anisotropy $\Delta = 
1/2$. 
We can then  interpret the 
model we studied as an integrable generalization  of the 
 deformed $SU(3)$ Perk-Schultz model  or the XXZ quantum chain at  anisotropy 
$\Delta = -(q +1/q) $ with $q=e^{i2\pi/3}$.

The BAE, whose solutions give the eigenspectrum of the model are quite 
difficult to solve analytically or numerically for general values of the 
parameter $t_p$ and density of particles (magnetization). Our results 
are summarized in Fig.~\ref{fig1}, where we have regions 1-5.

Regions 1,2,3 and 5 belong to a  critical phase governed by an 
underlying Coulomb gas conformal field 
theory with critical exponents varying continuously. 
We distinguished these regions according to the BAE roots of the low lying 
eigenenergies of the quantum chain. In regions 1 and 2 these roots are 
real. This fact allowed us to solve directly the BAE for quite large 
lattices ($L \sim 1000$).
 Exploring conformal invariance 
we obtain good estimates for the conformal anomaly and anomalous 
dimensions of operators of the underlying conformal field theory. 
In these two regions we could take the thermodynamic limit  and 
 obtained the critical exponents  in terms of 
integral equations (see \rf{H23}-\rf{H24}).

In regions 3,4 and 5 the 
BAE roots corresponding to the ground state are not real and very 
difficult  to calculate even for relatively small lattice sites.  In these 
regions our analysis were based on the direct calculation of the 
eigenspectra for lattices sizes $L\leq 24$.

Our results indicated that, regions 3 and 5, although having BAE complex roots 
in the ground state, have   the same critical behaviour as in regions  
1 and 2. Actually  regions 
 1,2,3 and 5 
are quite similar to the  XXZ quantum chain in the 
presence of a magnetic field, with anomalous dimensions $x_{l,m}$ given 
by  \rf{H22}, with the value of $x_p$ depending on $t_p$ and $\rho$.

As we cross from region 3 to region 4 there is a discontinuity of the 
ground-state energy. This is due to a crossing of the two lowest 
eigenenergies, and we expect that 
regions 3 and 4 are separated by a first order phase transition. 

Inside region 4 we found an oscillatory behaviour  for the energy gaps and 
momentum of the ground state, as we change the lattice size or  the 
 anisotropy parameter $t_p$. These oscillations are due to a large 
number of crossing of eigenlevels. These crossings made our 
finite-size analysis imprecise. We believe that probably such oscillatory 
behaviour is due to incommensurability effects on the 
charge (local magnetization) distribution in the lattice, as happens 
in other models whose incommensurability is established \cite{income}.

 We conclude mentioning two interesting open problems for the future. 
The derivation of the $R$-matrix associated to these new integrable 
spin one model and the extension of the present study to the second 
exact integrable model introduced in \cite{AB1}.
This second model at $t_p=1$ is related to the XXZ quantum chain with 
anisotropy $\Delta=-1/2$. This last model has remarkable properties.
 It is related to the problem 
of enumerating alternated sign matrices \cite{alternating} and the 
Hamiltonian with open boundaries is the evolution operator of a 
conformal invariant stochastic model, namely, the raise and peel model 
\cite{raise}. We then expect that the second model in \cite{AB1}  with values of $t_p \neq 1$ may also have 
interesting connections with other interesting problems in physics and 
combinatorics.

\section*{Acknowledgments}
We thank V. Rittenberg for discussions and a careful reading of the 
manuscript. This work has been partially supported by the Brazilian agencies FAPESP and 
CNPq.

\end{document}